\documentclass[twocolumn,aps,pra]{revtex4}
\usepackage{graphicx}
\usepackage{shortvrb,psfrag}
\begin{document}
\title{Deterministic Bell State Discrimination}
\author{Manu Gupta$^{1}$}
\email{er_manugupta@yahoo.com}
\author{Prasanta K. Panigrahi$^{2}$}
\email{prasanta@prl.ernet.in} \affiliation{$^1$ Jaypee Institute
of Information Technology, Noida, 201 307, India\\ $^2$ Physical
Research Laboratory, Navrangpura, Ahmedabad, 380 009, India}

\begin{abstract}
We make use of local operations with two ancilla bits to
deterministically distinguish all the four Bell states, without
affecting the quantum channel containing these Bell states.
\end{abstract}

\maketitle

Entangled states play a key role in the transmission and
processing of quantum information \cite{Niel,suter}. Using
entangled channel, an unknown state can be teleported \cite{bouw}
with local unitary operations, appropriate measurement and
classical communication; one can achieve entanglement swapping
through joint measurement on two entangled pairs \cite{pan1}.
Entanglement leads to increase in the capacity of the quantum
information channel, known as quantum dense coding \cite{Mattle}.
The bipartite, maximally entangled Bell states provide the most
transparent illustration of these aspects, although three particle
entangled states like GHZ and W states are beginning to be
employed for various purposes \cite{carvalho,hein}.

Making use of single qubit operations and the Controlled-NOT
gates, one can produce various entangled states in a quantum
network \cite{Niel}. It may be of interest to know the type of
entangled state that is present in a quantum network, at various
stages of quantum computation and cryptographic operations,
without disturbing these states. Nonorthogonal states are
impossible to discriminate with certainty \cite{wootters}. A
number of results have recently been established regarding
distinguishing various orthogonal Bell states
\cite{walgate,gosh1,vermani,chen}. It is counter intuitive to know
that, multipartite orthogonal states may not be discriminated with
only local operations and classical communications (LOCC)
\cite{walgate}. However, any two multipartite orthogonal states
can be unequivocally distinguished through LOCC \cite{vermani}. If
two copies belonging to the four orthogonal Bell states are
provided, LOCC can be used to distinguish them with certainty. It
is not possible to discriminate, either deterministically or
probabilistically the four Bell states, if only a single copy is
provided. It is also known that, any three Bell states cannot be
discriminated deterministically, if only LOCC is allowed.

Making use of only linear elements, it has been shown that a never
failing Bell measurement is impossible \cite{lutk}. A number of
theoretical and experimental results already exist in this area of
unambiguous state discrimination \cite{cola,pan2,kim}. Appropriate
unitary transforms and measurements, which transfer the Bell
states into disentangled basis states, can unambiguously identify
all the four Bell states \cite{pan2,kim,boschi}. However, in the
process of measurement the entangled state is vandalized. The
above is satisfactory, where the Bell state is not required
further in the quantum network.

We present in this letter, a scheme which discriminates all the
four Bell states deterministically and is able to preserve these
states for further use. As LOCC alone is insufficient for this
purpose, we will make use of two ancilla bits, along with the
entangled channels. Throughout the protocol, we will only employ
local unitary operations. At the end measurements are carried out
on the two ancilla bits, therefore we are able to preserve the
Bell states for further operations.

It should be noted that, the Bell states:
\begin{eqnarray} \nonumber
\label{bs}
|\psi^+\rangle = \frac{1}{\sqrt{2}}(|00\rangle+|11\rangle),\nonumber\\
|\psi^-\rangle = \frac{1}{\sqrt{2}}(|00\rangle-|11\rangle),\nonumber\\
|\phi^+\rangle = \frac{1}{\sqrt{2}}(|01\rangle+|10\rangle),\nonumber\\
|\phi^-\rangle = \frac{1}{\sqrt{2}}(|01\rangle-|10\rangle),
\end{eqnarray}
when operated on by single qubit operators such as Hadamard and
Pauli matrices, in solitude or in combination, get transformed
into each other. This is an interesting property which can easily
transform one Bell state to the other on demand. This property of
Bell states proves very handy in distinguishing them.
\begin{figure}
\begin{center}
\includegraphics[width=3in]{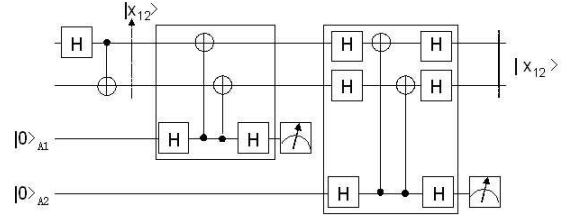}
\caption{\label{bsd}Diagram depicting the circuit for Bell state
discriminator.}
\end{center}
\end{figure}
Exploiting the above nature of the Bell states, we have designed a
circuit for the Bell state discrimination, as shown in
Fig.\ref{bsd}. It consists of two quantum channels, depicted as
$|x_{12}\rangle$, carrying the entangled state; two ancilla qubits
are used for carrying out local operations. In the end,
measurement is taken on these ancilla bits to know with certainty,
the type of Bell state that exists in the channel. Measurement on
the first ancilla will differentiate the four Bell states into two
pairs i.e., either $|\psi^+\rangle/|\phi^+\rangle$ or
$|\psi^-\rangle/|\phi^-\rangle$ as given in Eq.\ref{bst1}. While
the measurement on the second ancilla differentiates the Bell
states within these two groups as stated in Eq.\ref{bst2}. The
remarkable property of this circuit is that, the Bell states in
first two quantum channels retain their initial states, even after
being discriminated. Here, we have used Hadamard operation on the
entangled channel while differentiating the Bell states in
Eqs.\ref{bst1} and \ref{bst2}, though one can also use other
suitable single qubit operations. In table-I we have shown the
results of the measurements on both the ancillas when different
Bell states are present in the given circuit (Fig.\ref{bsd}).
Before measurement the states can be explicitly written as,

\begin{eqnarray} \label{bst1} \nonumber
|R_{A1}\rangle &=& [I_{2}\otimes I_{2} \otimes H]\ast[(x_1 \oplus
A_1) \otimes (x_2 \oplus A_1) \otimes I_2]\ast  \\ &&
[I_{2}\otimes I_{2}
\otimes H]\ast[|x_{12}\rangle \otimes |A_1\rangle] {\mathrm{~and}} \\
\label{bst2} \nonumber |R_{A2}\rangle &=& [H^{\otimes^3}]\ast[(x_1
\oplus A_2)\otimes (x_1 \oplus A_2) \otimes I_2]\ast \\ &&
[H^{\otimes^3}]\ast[|x_{12} \rangle \otimes |A_2 \rangle].
\end{eqnarray}
\begin{center}{\bf{Table. I:}}\end{center}
\begin{center}
\begin{tabular}{|c|c|c|}
\hline \hline
{\bf Bell State} & {\bf Measurement $A_1$} & {\bf Measurement $A_2$} \\
\hline
  $|\psi^+\rangle$ & $0$ & $0$ \\
  $|\psi^-\rangle$ & $1$ & $0$ \\
  $|\phi^+\rangle$ & $0$ & $1$ \\
  $|\phi^-\rangle$ & $1$ & $1$ \\ \hline
\end{tabular}
\end{center}

We acknowledge useful discussions with Prof. J. Pasupathy, S.
Dasgupta, J.N. Bandyopadhyay and A. Biswas.
%

\end{document}